# AUTHCOIN: VALIDATION AND AUTHENTICATION IN DECENTRALIZED NETWORKS

Benjamin Leiding, University of Göttingen, Germany, benjamin.leiding@cs.uni-goettingen.de

Clemens H. Cap, University of Rostock, Germany, clemens.cap@uni-rostock.de

Thomas Mundt, University of Rostock, Germany, thomas.mundt@uni-rostock.de

Samaneh Rashidibajgan, University of Rostock, Germany, samaneh.rashidibajgan@uni-rostock.de

## Abstract

*Authcoin is an alternative approach to the commonly used public key infrastructures such as central authorities and the PGP web of trust. It combines a challenge response-based validation and authentication process for domains, certificates, email accounts and public keys with the advantages of a block chain-based storage system. As a result, Authcoin does not suffer from the downsides of existing solutions and is much more resilient to sybil attacks.*

## 1 Introduction

Public key infrastructures (PKIs) are not only responsible for distributing and managing public keys, but also for ensuring a correct association between a public key and its owner. The hierarchical trust model for certificate authentication (commonly used by Certificate Authorities (CA) and web browsers) relies on hierarchically structured central authorities (Perlman, 1999), whereas the PGP[1] web of trust (WoT) uses a decentralized approach (Zimmerman, 1994). Instead of relying on central CAs, each user acts as an authority itself and ensures a number of bindings between (third) users and their public keys. Several security incidents in recent years have proven that CAs are vulnerable due to their centralized structure (Prins et al. 2011; Bugzilla, 2012; Espiner, 2012; Comodo Group, 2011). On the other hand, the decentralized PGP WoT does not provide sufficient certainty that the information stated in a public key is correct, since users do not carefully verify other users (missing incentives). In addition, it is trivial for malicious users to generate large numbers of key pairs and create structures which look like carefully verified keys without much effort.

Typically, PGP's WoT regards the following as criteria for trust: 1.) Number of signatures on the subject under investigation. 2.) Centrality of a node in the entire set of subjects. 3.) Timeline with information when signatures have been made (typically, signatures should appear during a longer period). 4.) Number of asymmetric trust relations within the WoT.

The basic idea is to increase the price for an attacker to pay for successfully manipulating the WoT. The attacker either has to pay real money or deliver a proof of work, which is a computing task that requires a sufficient amount of resources.

In this paper, we propose a new protocol called Authcoin which uses a flexible challenge-response schema for validation and authentication (V&A) of public keys. Depending on the chosen challenges, Authcoin can be used in various scenarios with customized security requirements and is not limited to be used in combination with PGP keys for email communication (as known from the PGP WoT) or website certificates (as for CAs). In addition, the protocol utilizes the transparent, fault tolerant, replicated and difficult to manipulate block chain (BC) concept (as known from Bitcoin (Nakamoto, 2008)) for storing data. As a result, Authcoin does not only provide a decentralized alternative to the commonly used PKI systems, but is also more resilient to sybil attacks than current decentralized approaches (as the PGP WoT).

---

[1] In the following, "PGP" will be used as a synonym for all PGP compatible software.

The paper starts with presenting the state of the art and describes commonly used types of PKIs. Section 3 provides a general description of the Authcoin protocol. Afterwards, Section 4 deals with potential threats to Authcoin, how to mitigate them and gives more detailed explanations on the challenge-response schema. Finally, Section 5 concludes the paper and Section 6 provides an outlook on future work.

## 2 State of the Art

Nowadays, there are two commonly used types of PKIs. The centralized system of CAs (Perlman, 1999) and its decentralized counterpart the PGP WoT (Zimmerman, 1994). Both have different advantages and disadvantages which will be described in the following section. In addition, a third approach, the Certcoin system (Fromknecht et al. 2014b) which has not been implemented yet, will be introduced.

### 2.1 Certificate Authorities

Certificate authorities are institutions or organizations inside a network which are treated as trustworthy by definition. They can sign individuals, organizations or another CA's certificates. Users who decide to trust a certain CA (and their decisions), also trust all individuals signed by this CA. The result is a tree-like, hierarchical structure with the initial CA (Root-CA) at the top of the system. This tree structure is also one of the biggest disadvantages of CAs since it introduces a single point of failure. The whole trust-system collapses as soon as a Root-CA gets compromised or untrustworthy for any reason. There have been several security incidents involving CAs in the last years. In 2011, an attacker issued certificates for domains of large IT-companies such as Google, Yahoo and others using an access to DigiNotar's (a Dutch CA) systems (Prins et al. 2011). As a consequence, DigiNotar's root certificate was removed from most browsers and the company went bankrupt. Another incident involved Trustwave Holdings. Trustwave operates a CA and issued a subordinate root certificate to a customer which enabled the customer to issue certificates on its own. The customer's identity was never revealed due to a Non-Disclosure Agreement (Bugzilla, 2012; Espiner, 2012). Another certificate incident of 2011 involved the CA Comodo (Comodo Group, 2011), where a compromised reseller account was used to issue arbitrary certificates.

Similar security problems might also occur in case a national authority forces a CA to cooperate and grant access to the CA's root certificates for surveillance reasons.

### 2.2 Web of Trust

In 1994 Phil Zimmerman described a decentralized counterpart to the centralized CA system: The PGP WoT (Zimmerman, 1994). Instead of relying on a central authority, each user acts as an authority itself and ensures a number of bindings between (third) users and their public keys. A successful verification of a public key results in an unidirectional signature between the public key of the verifier (Alice) and the verified user's (Bob) key. Such a signature is interpreted as a trust relation; Alice has successfully verified the authenticity of Bob's public key and therefore Alice decides to trust Bob's public key. Users can decide to trust a key if it is signed by somebody they trust or if there exists a chain of trusted signatures from their key to the targeted key. In contrast to the centralized CA system, the PGP WoT has no central point of failure. Nevertheless, it suffers from several other downsides such as missing incentives and a lack of punishments in order to motivate its users to adhere to the verification rules and contribute to the well being of the system.

It is also surprisingly trivial for malicious users to generate large numbers of keys and connect them in such a way, that the resulting network looks like a group of trustful users. Finally, as shown in a previous study (Leiding and Dähn, 2016), about 40 percent of the PGP WoT's email addresses are dead (not reachable any more) which makes us question the trustworthiness of signatures related to these unreachable email addresses.

As a conclusion, we can state that the WoT in its current form is worthless except in cases where it reflects direct trust relations.

## 2.3  Certcoin as an Alternative Approach

The Certcoin project (Fromknecht et al. 2014a; Fromknecht et al. 2014b) introduces an alternative approach and shares some basic ideas with Authcoin. Fromknecht et al. (2014a; 2014b) used "the consistency guarantees provided by cryptocurrencies such as Bitcoin and Namecoin (Namecoin, 2015) to build a PKI that ensures identity retention" (Fromknecht et al. 2014b). Certcoin does not require a central authority (similar to the PGP WoT) and uses the Bitcoin BC and the resulting advantages (decentralized, difficult to manipulate, distributed replication, fault tolerance, redundancy, transparency, etc.). The protocol provides methods for public key registration, update, revocation, recovery, verification and lookup. So far, Certcoin has not been implemented.

## 2.4  Our Contribution

Even though both, Certcoin and Authcoin, share the same underlying idea of a decentralized, BC-based WoT there are still major differences between them. Certcoin focuses on the combination of the PGP WoT and the BC concept as well as the feature of identity retention, but not on the V&A process itself or fighting malicious users. Authcoin combines the idea of a decentralized WoT, a fault tolerant, difficult to manipulate, replicated and transparent storage system (BC) and a (partially automated) bidirectional validation and authentication process. As a result, it is much more difficult for an adversary to introduce malicious key into the system or prevent the detection of such keys. Besides that, Authcoin is not limited to PGP keys, but can also be utilized for validation of domains, certificates, (email)-accounts or can be used in other similar scenarios.

# 3  Authcoin

The following section provides a general overview on Authcoin itself, its challenge-response-mechanism and further important concepts. Eventual security issues as well as advantages and disadvantages of Authcoin will be discussed later (Section 4) in order to keep the basic explanations as short as possible. As illustrated in Figure 1, we will first discuss the generation of a new key pair and how to establish an initial binding between the key and its owner (Section 3.1). Afterwards, the formal key validation (Section 3.2) as well as validation (Section 3.3) and authentication (Section 3.4) will be introduced. Later parts of this section deal with key revocation, expiration, recovery and lookup followed by a short overview on Authcoin's data storage (Section 3.8). Validation and authentication requests (VARs) will be introduced later in Section 4.3.

Even though the following sections use examples based on PGP key pairs, Authcoin is not limited to PGP and therefore most examples can also be applied to other public-key cryptography-based scenarios.

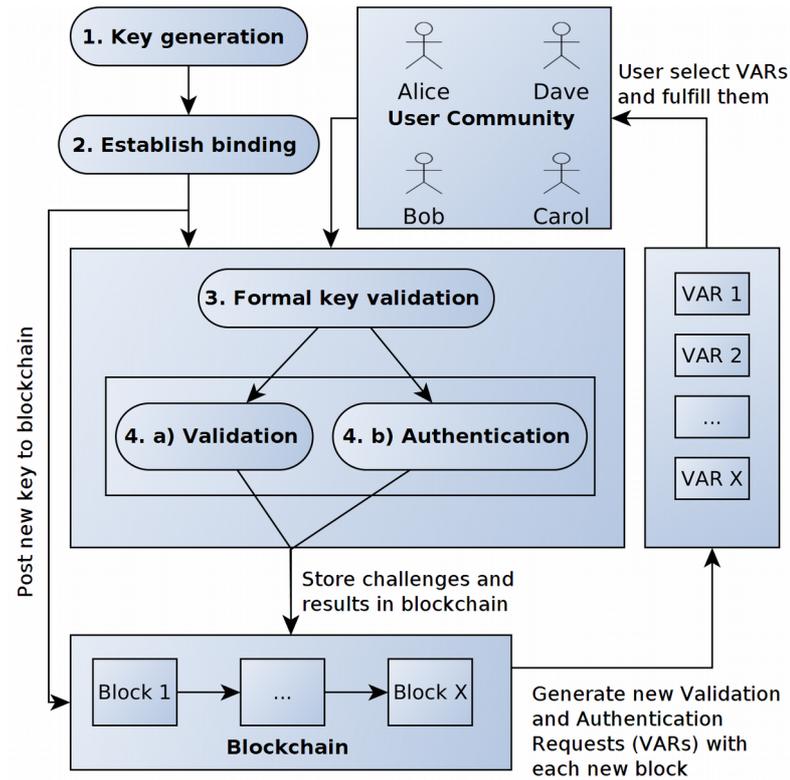

*Figure 1. General overview on Authcoin's workflow*

## 3.1   Generating a New Key Pair and Establishing an Initial Binding

As usual and known from the concept of the PGP WoT, each user generates a new key pair using a local PGP-compatible client and adds basic information (email, name, etc.) to the key pair. Establishing an initial binding between the generated key pair and its owner is crucial for authentication systems which rely on asymmetric cryptography. Traditional systems (e.g. PGP WoT) rely on (domain)-names and email addresses to identify users. Unfortunately, both of them are easy to fake and therefore it is trivial to create a public key for an arbitrary entity with a corresponding email address (e.g. obama@whitehouse.gov).

All accumulated information are collected and stored in the BC as described later in Section 3.8 and illustrated in Figure 1.

## 3.2   Formal Key Validation

Before the actual V&A, each involved public key is automatically checked for formal validity. The protocol validates the following properties (which are all stated in a common PGP public key): Is the key well formed (can PGP/GnuPG read the key as input)? Is the key length sufficient? Is the key still valid or already expired? Has the key been revoked?

In case all involved keys pass the formal validation, the actual V&A process starts. If necessary the mentioned example properties can be extended.

### 3.3 Domain, Certificate and (Email)-Account Validation

Authcoin's process of validating a domain and the key (certificate) of the domain is similar to the domain validation process deployed by "Let's Encrypt" (Let's Encrypt, 2015) as illustrated in Figure 2.

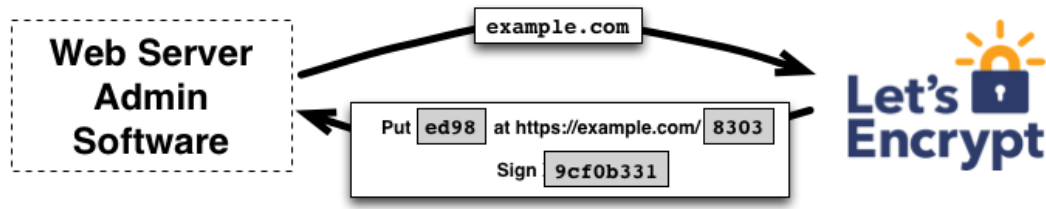

*Figure 2. Domain validation process as deployed by Let's Encrypt, Source: (Let's Encrypt, 2015)*

The domain owner runs the Let's Encrypt-Client on the domains machine, afterwards the client contacts the Let's Encrypt-Server (LES) and asks for a challenge. One challenge might be to provide a certain resource under a specific URI and sign it with the private key which corresponds to the validated public key which is validated. In Figure 2, the client is asked to provide the resource "ed98" at https://example.com/8303 and sign it with the private key. The client software fulfils the challenge as requested by the LES, which checks if the challenge's outcome is satisfying. If the validation succeeds, the domain owner has proven that he/she has access to the domain (domain validation), has access to the public and private key (key validation) and that the certificate (key pair) corresponds to the tested domain. In context of Authcoin, a similar validation process can be performed as illustrated in Figure 3. Since most keys inside the PGP WoT are used for encrypted email communication, the users are interested in a correct binding between a public key and the email address stated in the key, therefore we will use the validation of email accounts as an example.

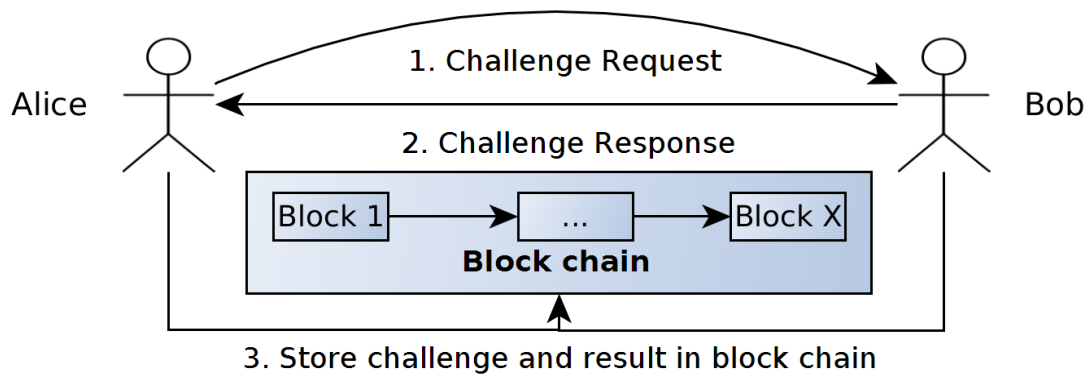

*Figure 3. General validation/authentication process as deployed by Authcoin*

As shown in Figure 3, Alice sends a challenge ("this is a challenge") encrypted with Bob's public key to Bob's email account. Bob is asked to fulfil the challenge, sign the response with his private key and send it back to Alice. Alice checks the results and can deduce (in case that the process finished successfully) the following three facts from the challenge and response: 1.) Bob has access to the email account (account validation). 2.) Bob has access to the public and private key (key validation). 3.) The key pair corresponds to the tested email account (binding).

The validation requests and results of the validation processes are stored as part of the BC (described in Section 3.8). Both, Alice and Bob, independently post the challenge and response to the BC.

It is important to keep in mind that neither in the domain/certificate validation example nor in the email account validation example the identity of the owner was verified, only that a certain entity actually had access to the domain or email account at the time the Authcoin protocol was run and that this entity had also access to the corresponding key pair. The authentication process is addressed later in Section 3.4.

An important difference between the domain validation process described earlier and the email account validation, is that this process is performed in both directions (bidirectional). Alice sends a challenge to Bob and receives (hopefully) a matching response and Bob does the same when receiving Alice's challenge in order to also verify Alice's mail account. As a result, it is more difficult for malicious users to introduce fake keys into the system and maintain introduced malicious keys. A major advantage of Authcoin's bidirectional validation processes is that they can be performed in an automated manner and even on large scale. As a result, each domain, certificate or account can be validated on a regular basis resulting in an improved overall security of the network.

### 3.4 Authentication

After validating the domains, certificates or email accounts, the problem of authentication is addressed. Similar to the validation procedures described before, Authcoin relies on a challenge-response-mechanism for authentication. In this section, we will use simple example challenges which might be used to verify the identity of an entity which is linked to a public key. More detailed explanations on challenges are provided later in Section 4.1.

Assuming the scenario that Alice and Bob went to the same school and have lost contact over the years, but somehow Alice managed to retrieve Bob's public key with an email address using Authcoin (it is also assumed that Alice has already performed the automated validation procedures as described earlier). Alice wants to verify Bob and his key in order to ensure that the email account and the public key belongs to Bob and not to somebody else. To do so, she sends a challenge to Bob and asks him to send her a picture of himself holding a copy of the current issue of a specific newspaper. In this simple example, it is plausible that Alice somehow knows what Bob might looks like and therefore is able to decide whether the person on the image is Bob or not. Since Bob also wants to make sure that he is really communicating with Alice, he verifies Alice in the same manner, resulting in a bidirectional verification of both participants. Again, as already mentioned for the validation procedure, both involved entities posts all information independently to the BC.

### 3.5 Revocations and Expirations

The following types of revocation and expiration are relevant for Authcoin:

#### 3.5.1 Key Revocation

Currently, Authcoin's key revocation is handled as known from PGP (Callas et al. 2007). A key is revoked by posting a key revocation certificate to the BC. In context of Authcoin, revoking a key is done by posting the key revocation certificate to the BC. Future versions of Authcoin might extend the protocol with a more sophisticated approach using a combination of offline and online key pairs, where the offline key pair can be used to revoke, update or replace the online key pair.

#### 3.5.2 Signature Revocation

Similar to the key revocation, a signature is revoked by adding a signature revocation certificate to the BC. A signature revocation expresses a total loss of trust in the signed key.

### 3.5.3 Key and Signature Expiration

Both, keys and signatures, have an expiration date. For security reasons, the lifetime of key pairs used with Authcoin is limited to a maximum of 12 months, afterwards a new key pair has to be created, but users can also decide to use shorter lifespans. Signatures either expire after a user-defined timespan (max. 12 months) or when the signing key or the key which got signed expire. An expired key cannot longer be used for V&A in context of Authcoin. Using the key outside Authcoin is still possible (even though it is not recommended).

## 3.6 Key Recovery

Currently, Authcoin does not support any key recovery mechanisms. Therefore, as known from the concept of the PGP WoT, a lost private key cannot be recovered. An alternative approach was introduced by Certcoin, which deploys a shared secret (Shamir, 1979; Blakely, 1979) solution. A user's private key secret is shared between a number of friends and at least two of them are required to restore the secret key (Fromknecht et al. 2014a; Fromknecht et al. 2014b). An advantage of this solution is the availability of a key recovery mechanism, but it comes with the downside of handling additional keys and the requirement of sufficient trusted persons. Furthermore, for non-technical users, the concept of public-key-cryptography alone is complicated enough; adding the concept of shared secrets demands to much from non-security experts.

## 3.7 Key Lookup

Authcoin's key lookup is done by traversing the BC and searching for a key described by one of its properties, such as the corresponding email address, a name, a key id, etc. Even though the BC might grow up to several gigabytes of data, traversing through the BC is not a real performance issue, since the lifespan of keys and signatures is limited to 12 months. Therefore, a key lookup only requires the traversal through data of the past 12 months.

## 3.8 Storing Information

Authcoin utilizes a BC-based transaction database (storage concept behind Bitcoin (Nakamoto, 2008), Ethereum (Buterin, 2014; Wood, 2014) and similar systems) as an underlying storage system which is used to keep track of keys, challenges, responses, signatures and all other relevant information. BC-based storage systems provide several desirable properties such as: decentralization (no trusted central authority), distribution of data, fault tolerance, transparency and redundancy. Furthermore, it is not possible to manipulate the BC as long as the majority of its users decide to do so.

A BC consists, as the name suggests, of an (technically) unlimited number of blocks which are chained together in a chronological order. Each block consists of transactions[2] which are the actual data to be stored in a BC. During the "mining"-process", miners collect valid transactions, compute a proof-of-work and the result is a new block which can be added to the BC. Each block depends on its predecessor block and therefore tampering and manipulating a block is quite difficult and requires an infeasible recalculation of all successor blocks.

For the ease of explanation, this paper assumes that Authcoin has its own, independent BC. But it is also possible to utilize existing BCs for this purpose as done by Namecoin which is built on top of Bitcoin and uses the same BC. Alternative projects as Ethereum (Buterin, 2014; Wood, 2014) maintain their own, independent and customized chain.

---

[2]In case of Authcoin, transactions are the stored data records such as challenges, responses, etc.

# 4 Security and Reliability

Section 3 provided only a general overview on Authcoin and did not discuss any security issues or attacks on the system. The following section deals with exactly these topics, how to mitigate possible attacks, prevent misuse of Authcoin and explains Authcoin's challenge-response concept in detail.

## 4.1 Challenges

Security and reliability of Authcoin heavily depend on the challenges chosen for the V&A of keys and identities. Therefore, this part gives more detailed information on the challenge-response-concept and related security issues as well as a few example challenges. Again, PGP and email-based communication are used as an example for the ease of explanation even though Authcoin is not limited to this scenario.

### 4.1.1 Basic Challenges

In Section 3.3 as well as in Section 3.4, we discussed a special subtype of V&A (called local V&A) were two entities either had or have at some point the possibility of direct personal interaction (e.g. meeting each other in person). V&As without such an opportunity where it is not possible that the two entities interact directly with each other, are called global V&As. In general we have to distinguish three basic types of V&A: Local V&A with additional information (e.g. a side channel), global V&A with additional information (e.g. a side channel) and global V&A without any additional information.

The easiest and most convenient category is a local V&A with additional information. Two examples were provided in Section 3.3 and Section 3.4. Both participating entities had additional knowledge about the other entity such as personal (old) information (name, parents, current/former address, date of birth, etc.) or even an existing side-channel for exchanging information (e.g. alumni meetings). Therefore, it is relatively easy to construct a personalized challenge-based on shared experiences or information which is (in a best case) only known to both entities. The most common example of the PGP WoT: "Two entities meet in person, verify each others key and prove their identity using their identity cards or passports" belongs to this category as well as requesting an image of the passport owner with its passport. In this case, the identity card or the passport is the additional information and both parties have decided to trust them as an authentication method (passports can be faked (e.g. by the government itself) or stolen).

A bit more challenging to handle are global V&As with additional information or a side-channel. An example scenario might look like this: Alice is a well known blogger and Bob a reader who wants to communicate with Alice for some reason. Unfortunately, Alice and Bob do not know each other and do not have any common acquaintances. Bob found a public key with Alice's email address and her name in the Authcoin system. He wants to make sure that the key really belongs to Alice and therefore creates a challenge: "Alice if this your key, please use the following keyword X in your next blog post". The challenge is sent to the email address stated in the key and posted to the BC. Alice fulfils the challenge by using the keyword in her next blog post and sends the response to Bob. Bob reads the post and publishes it. For the local V&A with additional information as well as the global V&A with additional information, the security of the challenge highly depends on the security (e.g. tamper proof) of the side-channel or the fact that the additional information is only known to the verifier and the verification target but not to someone else.

Much more difficult are global V&As without any additional information. Bob wants to communicate with Alice and found a public key of Alice in Authcoin's BC. Unfortunately, Bob has no additional information about Alice (Image, etc.) and no side-channel as in the previous example. In this case, Authcoin can only provide validation but no authentication (at least with the current set of challenges, but this might change in the future). Using the process described in Section 3.3, it is still possible to validate that the entity behind the public key has access to the specified email account and the corresponding private key, but it is not possible to determine whether this entity is Alice or not. Nonetheless, Alice and Bob can decide to give up some security guarantees and decide to use a non-verified additional information such as an identity card. In this case, Bob has no way to either check whether the image on the ID belongs to Alice (since he does not know what Alice looks like), nor verify that the ID is not faked (maybe even by the government to fool Bob) or stolen. But since Authcoin stores all challenges and responses in the BC, potential malicious users risk exposure through other genuine users while analysing the BC.

It is important to note that for the three V&A types discussed above a signature between two keys is only created in case that the validation process (validation signature) or authentication process (authentication signature) was successful. A failed validation or authentication is also documented in the BC. By not creating signatures for failed validations or authentications, Authcoin stays compatible for already existing PGP systems without introducing new signature types (the distinction between V&A signature is an internal attribute of Authcoin). In contrast to current authentication processes such as known from the PGP WoT, Authcoin requires most validations and authentications to be performed in both directions (bidirectional) instead of unidirectional. Bidirectional V&A results in more frequent examination of each key, identity, domain or certificate and makes it more difficult for malicious entities to stay undetected (more on malicious entities in Section 4.2). Another advantage of Authcoin V&A approach is a lower threshold for users to participate in V&A since it is not necessarily required to meet the verification target in person (even though this is still possible and just a different type of challenge). Finally, automating the validation process provides each user with the unique possibility to validate all existing entities of the Authcoin-system on its own without relying on any transitive relations (nevertheless it is possible for authentications if desired by the user).

As already shown through the mentioned examples above, Authcoin is not fixed to a specific type of challenge. Instead it is meant to be as flexible and extensible as possible and use all kind of challenges. Therefore, the results of future research on challenges can easily be integrated in Authcoin, especially new challenges which are more secure and harder to manipulate than the existing ones. Another advantage of this approach is that Authcoin can also be deployed in other scenarios as the ones described above, for example in IoT (Internet of Things) environments.

### 4.1.2 Protecting Privacy

Storing challenges and responses in a transparent and publicly available BC might result in some issues regarding to the users privacy. In order to address this issue, Authcoin supports a special type of private challenge. In this case, the challenge and response are stored in an encrypted manner in the BC and the keys are only known to the involved parties. To prevent abuse of this opaque privacy enhanced challenge, Authcoin does not create a signature for such successfully passed challenges and only stores the created challenge and response as part of the BC. Thereby it is still possible for other users to see that a successful V&A took place, but malicious users cannot create signatures based on opaque challenges which cannot be verified by others. This transparency property is an important feature of Authcoin since it also allows uninvolved users to track the origins of a signature.

### 4.1.3 Adaptable Level of Required Security

As mentioned earlier, the challenges influence the overall security and reliability of the system. As a result, adapting the requirements for the deployed challenges leads to a different level of provided security. In some scenarios, deploying only validation mechanisms might be enough for a given purpose. In other scenarios, it might be necessary to combine different challenges based on different identifiers in order to provide a maximum level of security and reliability. Many other scenarios lie in between these two extrema. For example in some cases, a picture with the verification target and its passport (including picture, etc.) provides a reasonable level of security, in other cases a passport does not proof anything since possible adversaries can create any desired passport identity.

A further security improvement might be to utilize biometric identifiers which are more difficult to fake. Commonly used biometric identifiers are fingerprints, eyes (retina or iris recognition), voice, face (face recognition) or DNA. Biometric identifiers can either be used to derive a new key pair from the identifier (Bjorn, 2000) or get included into the new key pair in addition to the traditional identifiers. Adding biometric identifier to the key pair establishes additional bindings between the key owner and the key pair.

It is important to keep in mind that users cannot solely rely on successfully passed challenges. They have to analyse and examine the challenges used during the V&A process of a key and have to decide on their own if these challenges provide a suitable level of reliability and security or not.

## 4.2 Malicious Entities

As known from our experiences with the PGP WoT, the biggest security issue of a decentralized trust system are malicious users (e.g. sybil nodes). Authcoin implements several restrictions and mitigations concepts in order to prevent malicious users from damaging the system and identifying them as soon as possible. Similar to the PGP WoT and comparable solutions, Authcoin cannot prevent (at least not at the moment) the participation of sybil nodes. Nevertheless, due to the adaptable challenge-response mechanism and the transparent storage of all information, it is much more difficult for an adversary to keep its malicious nodes undetected.

The first line of defence is the challenge-response-mechanism which can (if used correctly) detect and identify sybil nodes since they might not be able to successfully pass the proposed challenges. Making challenges as tamper-proof as possible makes it much more difficult for sybil nodes to stay undetected. For example, the verifier might not only ask for an image of the verification target with its passport, but also with an apple in his/her right hand and his/her left eye closed. Adding further constraints makes it more difficult for an adversary to use existing images of the verification target and manipulate them according to the challenge especially under time restrictions. In addition, a list of several constraints makes it also significantly more difficult to create fictional users (with fictional computer generated avatars) in short time, which fulfil the challenge in a satisfying manner. Using images of existing persons from the Internet in combination with arbitrary names is also dangerous for an attacker since the images are stored in the publicly available BC and might be discovered by other users which know the person on the image and her/his real name. Figure 4 illustrates how users can detect such a mismatch. Furthermore, future research on applicable challenges might lead to more secure and tamper-proof challenges which make it nearly impossible for sybil nodes to stay undetected.

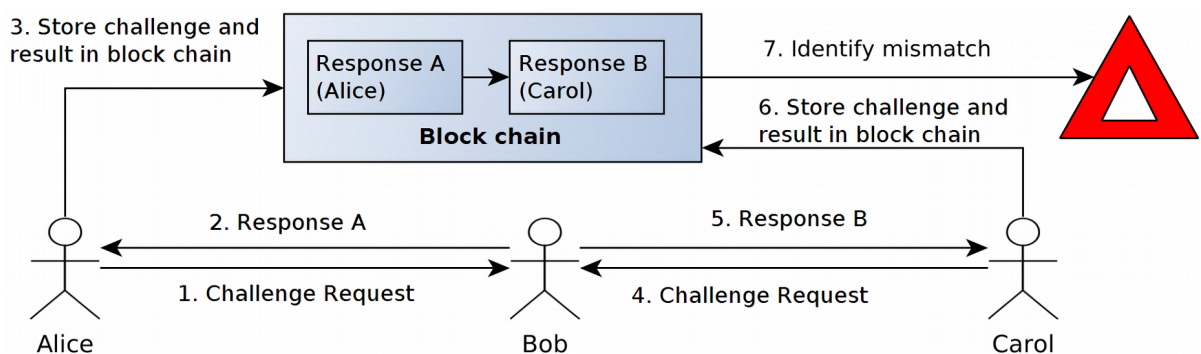

*Figure 4. Identify mismatch in the block chain*

Deploying a mandatory bidirectional authentication process is an additional burden for malicious users. It might be still possible to create a collective of sybil nodes with signatures between the participating entities, but as soon as nodes outside this collective interact with the collective the probability of exposure increases.

As a result of limiting keys lifespan to a maximum of 12 months, maintaining such sybil collectives is much more time consuming than before. In case of the PGP WoT, it is trivial to create an arbitrary number of keys with unlimited life span, connect them among each other and then create outgoing signatures to legitimate nodes and also might receive some back signatures from unreliable verifiers, which finally results in a permanent incorporation of the sybil collective in the "web of trust". Moreover, the transparent nature of Authcoin makes it also easier to identify unreliable verifiers which do not take the V&A process seriously and identify them (and their actions) as not trustworthy.

Another approach for detecting and mitigating malicious nodes are automated V&A requests which will be introduced in the following section in more detail.

### 4.3 Automated Validation and Authentication Requests

Automated validation and authentication requests (VARs) are automatically and randomly created during the mining process (more details on mining in Nakamoto (2008)) of a new block of the BC (as illustrated earlier in Figure 1). The number of generated VARs depends on the number of existing and still valid (not expired, revoked, etc.) keys in the system. An automated VAR expresses the desire of the system to validate and/or authenticate a randomly chosen entity inside the system. VARs are publicly stored as a part of the BC and can be fulfilled by Authcoin's users. Deploying such an automated requests-mechanism results in several benefits compared to existing solutions:

Firstly, the approach of automated VARs makes it much easier to break into sybil collectives "by accident" and expose them as such in case they fail the validation or authentication process. Identifying one sybil node leads to questioning all other nodes which claim to have successfully validated and/or authenticated the sybil node and identifying these nodes either also as sybil nodes or at least as unreliable verifiers. Due to VARs and the bidirectionality of authentications, it is also possible to increase the number of V&As for each key, resulting in higher probabilities of detecting malicious users.

Especially in context of certificate and email account validation, VARs result in interesting application scenarios: Issuing large amounts of validation requests and storing the results in the BC, results in a large regularly updated dataset containing information about the binding between a certificate/key and a website/email account over time. Therefore it is possible to track which certificate/key was used at which date and time for which website/email account. Even more important in context of the Snowden revelations and highly sophisticated targeted attacks: It is possible to view certificates/keys (illustrated in Figure 5) from different points of view; can Alice and Bob "see" the same certificate for a given website even though Alice accesses the website from the US and Bob from China (Great Firewall)?

Is this the case for all users from China or only for Bob? Does the certificate match the website? Combining this approach with a browser-plugin results in an interesting security feature even for the average Internet users and not only security experts. Especially since the fulfilment of the validations can be done automatically by a client on the user's devices without any interaction from the users. Moreover, using email account validation, it is also possible to keep track which email addresses are still technically reachable and operated and which are already dead. As shown by Leiding and Dähn (2016), about 40 percent of the PGP WoT's email addresses (associated with the PGP keys) are technically not reachable any more without anybody noticing. By using Authcoin's validation procedure, it is possible to monitor which addresses are still technically reachable (does not mean that anybody uses them any more).

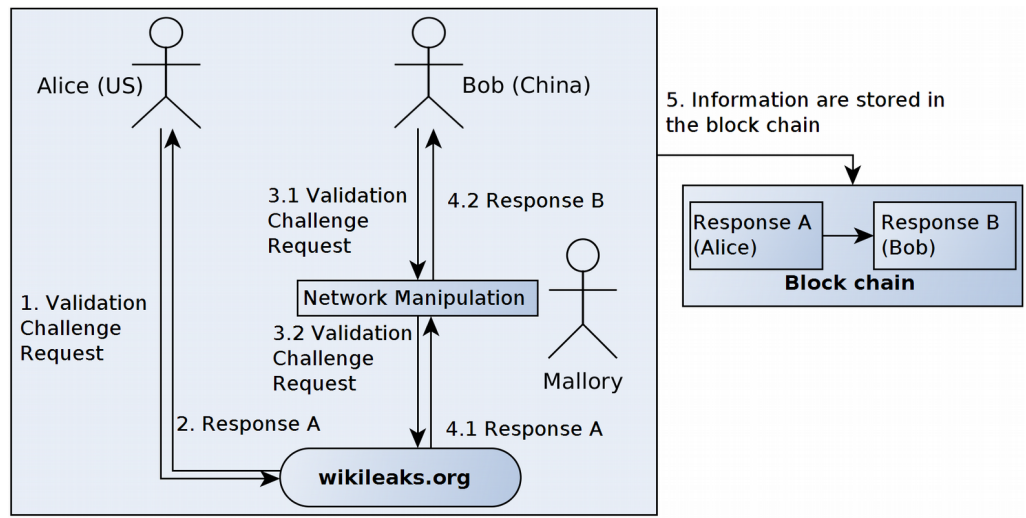

*Figure 5. Identifying certificate manipulations through Authcoin*

In order to make it more difficult to misuse the VAR-mechanism, we introduced some restrictions: For example, VARs can not be issued by users manually, instead they are generated automatically after a new block was added to the BC. If malicious users could create any desired number of requests on their own (and fulfil them afterwards with keys under their control), the relevance of the automatically generated requests would be almost zero. In order to avoid similar tactics for the automatically generated VARs, a hash-based selection algorithm decides whether a specific user is allowed to fulfil the VAR or not. The selection algorithm simply calculates the hash of the concatenation of the VAR and the (possible) verifier's key. If the binary presentation of the result starts with a certain combination (e.g. a 1 or a 0; 10 or 11, etc.) the user is allowed to fulfil the VAR, otherwise not. This decision seems to be random, but that is exactly the idea behind it. The algorithm is used to make it more difficult for malicious users to fulfil VARs with other keys under their control. Of course this approach can be undermined, but it increases an attackers cost of not getting exposed. Besides limitations through the selection-algorithm, it is also necessary that the key used to fulfil the VAR itself was created before the VAR in order to avoid that attackers create keys to fulfil the VAR after they discovered it.

In future versions of Authcoin, the VAR-mechanism might be combined with an incentive system in order to encourage users to fulfil VARs on a regular base (in case that is even necessary; besides that, fulfilling validations does not require the user's interaction at all). The incentive system might be part of an overall trust metric concept which not only includes the results of V&As, but also rewards behaviour that benefits the system (such as fulfilling VARs).

## 4.4 Bootstrapping

The bootstrapping process of Authcoin turns out to be rather simple. It does not matter if the first users of Authcoin are genuine or not. In the long-term, the genuine users of the protocol will figure that out and mark these three as not trustworthy and post the information to the BC.

## 4.5 Possible Attacks on Authcoin

As Authcoin uses a similar method for domain authentication as "Let's Encrypt", it is vulnerable to the same attack vector. It is possible that an attacker runs an attack on the routing and/or DNS system to the effect that for a short period of time the communication between the Let's Encrypt server and the machine of the domain in reality takes place between the Lets Encrypt server (LES) and a machine of the attacker. The machine of the attacker contacts the LES and asks for a challenge. The LES provides such a challenge. The attacker receives the challenge, signs it with its own private key and provides it under the proposed URL. When the LES checks the response to the challenge, again the LES is directed to the machine of the attacker. As a result the verification of the challenge by the LES turns out to be valid (with respect to the public key of the attacker). Thus, the LES provides the attacker with a certificate where the public key of the attacker is certified to be the public key of the original impersonated domain.

A requirement for a success of this attack is that the attacker can manipulate the network in such a manner that, for a short period of time, the LES believes to be communicating with the original domain but instead communicates with the attacker. Admittedly, such an attack is difficult. However, if such an attack would be impossible we would not need certificates for server authentication after all.

The protection offered by the Let's Encrypt approach, however, easily can be improved: If several validating servers provide challenges, the attacker must impersonate all these challenges. This improvement comes at the difficulty that any of the LES could launch a DoS attack against a user by claiming that the response did not correctly match the challenge. In case of Authcoin, such an unjustified claim can easily be detected since both parties (verifier and the verification target) post the challenge and responses independently to the BC. So as a result, under certain very difficult to achieve circumstances it is possible to attack the Authcoin protocol, but since this an attack on very fundamental parts of the Internet's infrastructure it cannot be prevented by Authcoin.

# 5 Conclusion

Current public key infrastructure solutions come along with several downsides and disadvantages resulting in the need of a new approach which fixes the existing problems. Authcoin aims to do that in a decentralized and distributed manner using a BC-based storage system and a redefined challenge-response-based V&A process for public keys, certificates and email accounts. Due to its transparent nature and public availability, it is possible to track the whole V&A history of each entity in the Authcoin system which makes it much more difficult to introduce sybil nodes and prevent such nodes from getting detected by genuine users. In addition, Authcoin provides some additional features such as automated VARs and a certificate/email account monitoring system which enhances the overall security. Finally, due to the flexible challenge-response-scheme, the level of security provided by Authcoin can be easily adapted and allows Authcoin to be deployed in various application scenarios.

# 6 Future Work

Besides implementing and testing Authcoin under real world conditions, future work might also focus on leveraging the concepts of the V&A process. There is no necessity for key pairs to be part of the V&A process. Instead, Authcoin could be generalized and used for all kind of V&A.

In addition, Authcoin's security might be improved through the development of new, more tamper-proof challenges which might include biometric identifiers such as fingerprints and voice-samples or iris and retina information. Furthermore, Authcoin might benefit from deploying trust metrics on top of it using the data records which are part of the BC in order to provide an estimation of a keys trustworthiness.